\documentclass[nofootinbib,amsmath]{revtex4}

\newcommand{\beq}{\begin{equation}}
\newcommand{\eeq}{\end{equation}}

\newcommand{\beqa}{\begin{eqnarray}}
\newcommand{\eeqa}{\end{eqnarray}}

\newcommand{\bean}{\begin{eqnarray*}}
\newcommand{\eean}{\end{eqnarray*}}
\newcommand{\ra}{\rightarrow}

\newcommand{\al}{\alpha}

\usepackage{graphicx}
\usepackage{citesort}
\usepackage{mathbbol}

\begin{document}

\title{Foliations and 2+1 Causal Dynamical Triangulation Models}

\author{Tomasz Konopka} 
\email{tkonopka@perimeterinstitute.ca} \affiliation{University of 
Waterloo, Waterloo, ON N2L 3G1, Canada, and \\ Perimeter 
Institute for Theoretical Physics, Waterloo, ON N2L 2Y5, Canada}
 
\begin{abstract}
The original models of causal dynamical triangulations construct 
space-time by arranging a set of simplices in layers separated by 
a fixed time-like distance. The importance of the foliation 
structure in the $2+1$ dimensional model is studied by 
considering variations in which this property is relaxed. It 
turns out that the fixed-lapse condition can be equivalently 
replaced by a set of global constraints that have geometrical 
interpretation. On the other hand, the introduction of new types 
of simplices that puncture the foliating sheets in general leads 
to different low-energy behavior compared to the original model.
\end{abstract}
\maketitle

\section{Introduction}

Causal dynamical triangulation models are a non-perturbative 
approach to the study of quantum gravity. In recent years, these 
models have been used to construct a Lorentzian path integral for 
quantum gravity as a sum over geometries constructed by gluing  
many primary building blocks together \cite{PathIntegral}. A 
$1+1$ dimensional model has been fully solved analytically 
\cite{SolveEaxactly} and higher dimensional versions have been 
studied numerically \cite{Simulations, TriangLorentz, 
FourDWorld}. Simulations in $3+1$ dimensions suggest that causal 
dynamical triangulations generate large-scale space-times with 
desirable properties \cite{FourDWorld}. 

Causal dynamical triangulations are based on the Regge action for 
simplicial gravity, where macroscopic space-times are constructed 
from elementary building blocks glued together. These elementary 
building blocks are $n-$simplices, with $n$ ranging from zero to 
the dimensionality of the space-time. The Lorentzian nature of 
the space-time is implemented by making some of the edges in the 
simplices time-like. In the original models, the elementary 
building blocks come only in a few varieties 
\cite{TriangLorentz}. The constructions give rise to foliated 
space-times in which every layer of simplices is separated from 
another layer by a space-like hypersurface formed by the faces of 
the simplices. Hence, there is a clear distinction between 
space-like and time-like directions. 

The foliation structure has been argued to play an important role 
in giving the Lorentzian models desirable low-energy properties 
such as an appropriate Hausdorff dimension \cite{TriangLorentz}. 
Given that the same properties do not arise in Euclidean models, 
it is interesting to ask how rigid the structure of the studied 
Lorentzian models really is. What are the consequences of 
loosening the assumptions in the original Lorentzian models, and 
is it possible to build more general models with the same 
low-energy properties? This paper address two aspects of this 
question.

One important aspect of the original Lorentzian model is the 
requirement for all time-like edges to have equal length-squared, 
which can be thought of as a `lapse' constraint. In $1+1$ 
dimensions, this requirement on the individual simplices can be 
removed so that their shapes and sizes can be chosen quite freely 
as long as a global constraint on the ensemble of simplices 
making up the entire space-time is enforced instead 
\cite{FotiniLee}. One of the results of this paper is that a 
similar generalization is possible in higher dimensions, where 
the analogous global constraints can be interpreted as specifying 
average volume and/or curvature contributions for the simplices 
in the triangulation.

A second aspect of the original model is connected with the types 
of simplices that are used as the primary building blocks of 
space-time. The existence of global hyper-surfaces constructed 
from the space-like faces of simplices can be compromised by 
introducing new types of simplices into the triangulation. 
Introducing new types of simplices that puncture the 
hyper-surfaces foliating the space-time in general causes 
triangulation models to move into a different equivalence class 
than that of the original models. Therefore, the extended models 
may have different low energy properties than the original ones. 
The new models, however, can reproduce the familiar behavior if 
additional but ad-hoc constraints are introduced to restrict the 
number of foliation-puncturing simplices that appear in the 
triangulations.

This paper focuses on $2+1$ dimensional models for concreteness 
but, where possible, the discussion is extended to higher 
dimensions. Section \ref{s_model} is a short summary of the 
original Lorentzian model in $2+1$ dimensions and includes a 
discussion of its associated partition function. Section 
\ref{s_lapse} discusses how the lapse can be allowed to vary. The 
following section \ref{s_broken} deals with the new types of 
simplices that puncture the foliation hypersurfaces. A summary of 
the findings appears in section \ref{s_conc}.

%%%%%%%%%%%%%%%%%%%%%%%%%%%%%%%%%%%%%%%%%%%%%%%%%%%%%%%%%%%%%%%%%%%%%%%%%%%%%%%%%%%%%%%%%%
\section{Original Model \label{s_model}}

\begin{figure}
\begin{center}
  \includegraphics[scale=1]{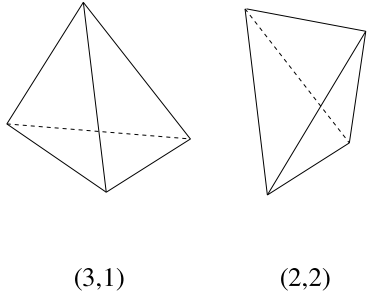}
  \end{center}
\caption{ \label{f_old} Simplices in 2+1 dimensions. Edges 
pointing mostly vertically (mostly sideways) are time-like 
(space-like) and have negative (positive) edge-lengths. Simplices 
are named after the number of vertices on space-like surfaces. }
\end{figure}

In Lorentzian dynamical triangulation models in $2+1$ dimensions, 
a set of tetrahedra are glued together along their faces to 
construct an extended space-time. The original model uses the 
$(3,1)$ and $(2,2)$ types of tetrahedra shown in figure 
\ref{f_old}. This section reviews the original $2+1$ model; for 
more details, see \cite{TriangLorentz}. 

In the original model, all space-like edges have length-squared 
set to unity, and all time-like edges have length-squared equal 
to $-\alpha$, where $\alpha$ is a positive constant. Space-times 
formed by gluing these simplices together always form a layered 
structure and contain space-like surfaces formed by the 
space-like faces of the $(3,1)$ tetrahedra. These surfaces can be 
thought of as the discrete analogs of foliation hyper-surfaces.
The Regge action corresponding to such a configuration is \beq 
\label{Sgeneral} S = k_0 \sum_a \left(-i V(a) \left(2\pi - \sum 
\theta\right) \right) +
    k_0 \sum_b \left(V(b)\left(2\pi-\sum\theta\right)\right) -
    \lambda \sum_{c} V(c),
\eeq where the indeces $a$, $b$ and $c$ run over all space-like 
edges, time-like edges, and $3-$simplices in the triangulation, 
respectively. $k_0$ is related to Newton's constant and $\lambda$ 
is related to the cosmological constant. The real functions $V()$ 
denote the lengths of edges or volumes of tetrahedra, and the 
sums of $\theta$'s tally the dihedral angles subtended by faces 
meeting at an edge. Inserting explicit expressions for the angles 
and volumes associated with the simplices of figure \ref{f_old}, 
the action can be written as \beqa \label{S3full} S &=& k_0 
\left( \frac{2\pi}{i}N_1^S -
  \frac{2}{i}N_3^{(2,2)}\sin^{-1}\frac{-i2\sqrt{2}\sqrt{2\alpha+1}}{4\alpha+1} -
  \frac{3}{i}N_3^{(3,1)}\cos^{-1}\frac{-i}{\sqrt{3}\sqrt{4\alpha+1}} \right) 
\nonumber \\
&&+k_0 \sqrt{\alpha} \left( 2\pi N_1^T -
   4N_3^{(2,2)}\cos^{-1}\frac{-1}{4\alpha+1} -
   3N_3^{(3,1)}\cos^{-1}\frac{2\alpha+1}{4\alpha+1}\right)  \\
&&-\lambda\left( 
N_3^{(2,2)}\frac{1}{12}\sqrt{4\alpha+2}+N_3^{(3,1)} 
\frac{1}{12}\sqrt{3\alpha+1}\right). \nonumber \eeqa The 
variables $N_1^S$, $N_1^T$, $N_3^{(3,1)}$ and $N_3^{(2,2)}$ count 
the number of space-like edges, time-like edges, and 
$3-$simplicies of the two kinds, respectively. It is possible to 
manipulate this expression in order to remove the dependence on 
the number of edges and tetrahedra of each allowed kind from this 
expression. This leaves $S$ as a function of $N_1=N_1^T+N_1^S$ 
and $N_3=N_3^{(3,1)}+N_3^{(2,2)}$ only.

The model is formulated with Lorentzian signature, but it is 
possible to perform a Wick rotation into Euclidean space by 
changing the sign of $\alpha$. After setting $\alpha=-1$, the 
action becomes \beq S\ra iS_E; \qquad\qquad S_E = k_3 N_3 -k_1N_1 
\eeq where $k_1$ and $k_3$ are \beq \label{k1k3defs} k_1 = 2\pi 
k_0, \qquad\qquad k_3 = k_A k_0 + \lambda k_V = \left(6 
\cos^{-1}\frac{1}{3}\right)k_0 + \lambda\frac{1}{6\sqrt{2}}. \eeq 
The final equation serves as a definition for $k_A$ and $k_V$. 
The Euclidean action can also be rewritten as \beq 
\label{canonicalaction} S_E = k_V N_3 (\lambda - k_0\zeta) \eeq 
where \beq \label{xidef} \zeta = \frac{k_1}{k_0 k_V} \xi - 
\frac{k_A}{ k_V}, \qquad\qquad \xi = \frac{N_1}{N_3}. \eeq 

The action is traditionally discussed in terms of the 
dimensionless parameter $\xi$. This parameter is known to take a 
maximal value of $4/3$ in purely Euclidean models, but, as is 
shown below, can only be as high as $5/4$ in the Lorentzian 
version discussed here. Since numerical simulations show that the 
Euclidean and Lorentzian models have very different properties, 
one can treat $\xi$ as a suggestive indicator for the large-scale 
behavior of a model. In a vague sense, $\xi$ can be thought of as 
measuring the connectivity between simplices. The other parameter 
$\zeta$, obtained from $\xi$ by a re-scaling and a shift, is 
introduced for future convenience. 

The remaining part of this section is dedicated to estimating the 
allowed ranges for $\zeta$ and $\xi$ and sketching how these 
ranges affect the partition function of the triangulation model 
in the large volume limit. To estimate these ranges, one can 
write down an arbitrary configuration of tetrahedra and define a 
set of moves, i.e. manipulations of the triangulation, that allow 
to generate other configurations starting from the initial one. 
The minimal and maximal values of $\xi$ can be estimated from the 
way these moves change the configuration of simplices.

It is convenient to keep track of the configuration via a vector 
$f$ counting the number of simplices of various dimensions, \beq 
f = f(N_0, N_1^S, N_1^T, N_3). \eeq Here $N_3$ is the sum of 
$(3,1)$ and $(2,2)$ types of $3-$simplices. It is believed that 
just three types of moves are sufficient to manipulate an initial 
configuration of simplices into any other one 
\cite{TriangLorentz}. They are given names like $(a\ra b)$, 
indicating that they change a configuration of $a$ tetrahedra 
into another configuration with $b$ tetrahedra. For example, one 
kind of move replaces two $(3,1)$ tetrahedra glued along a common 
space-like face by a group of six $(3,1)$ simplices (see 
\cite{TriangLorentz} for details.) The way this and another 
$(2\ra 3)$ move change the $f$ vector is described by \beqa
\Delta_{(2\ra6)}f &=& (1,\,3,\,2,\,4), \nonumber\\
\Delta_{(2\ra3)}f &=& (0,\,0,\,1,\,1). \label{oldmoves} \eeqa 

%\Delta_{(4\ra4)}f &=& (0,\,0,\,0,\,0,\,0,\,0),  \\

Starting with a triangulation consisting of $N_1$ edges and $N_3$ 
tetrahedra, one can construct another configuration by performing 
$x$ moves of the $(2\ra 6)$ kind and $z$ moves of the $(2\ra 3)$ 
kind. The total change in the $f$ vector in such a transaction 
would be \beq \Delta f = (x,\,3x,\,2x+z,\,4x+z), \eeq giving a 
resulting ratio $\xi$ of \beq \label{xi3temp} \xi = 
\frac{N_1+5x+z}{N_3+4x+z}. \eeq In the limit of large $x$ and 
$z$, which corresponds to constructing triangulations of large 
volume, \beq \lim_{x\ra \infty} \xi = \frac{5}{4}, \qquad\qquad 
\lim_{z\ra \infty} \xi = 1. \eeq These are in fact the upper and 
lower bounds for the parameter $\xi$ after many substitutions, so 
the range for $\xi$ is \beq \label{xibound} 1 \leq \xi \leq 
\frac{5}{4}. \eeq Using (\ref{k1k3defs}) and (\ref{xidef}), this 
can be translated into \beq \label{zetabound} -9.356 < \zeta < 
3.973. \eeq

The range of $\zeta$ plays an important role in determining the 
partition function of the triangulation model \cite{Gabrielli}. 
The partition function is \beq \label{Z1} Z(k_0,N_3) = \sum_{T} 
W(k_0,N_3,T) e^{-k_V N_3(\lambda - k_0\zeta)} \eeq where 
$W(k_0,N_3,T)$ is a weighting function that depends on the 
symmetry of a configuration $T$ and the sum is over all possible 
configurations having a fixed total number of simplices $N_3$. In 
the limit of a large fixed volume $k_V N_3$, the expression for 
the partition function can be considerably simplified. Up to a 
factor, $Z$ can be written as \beq \label{Z2} Z(k_0,N_3) \sim 
\sum_{T} W(k_0,N_3,T)e^{k_V N_3 k_0 \zeta}.\eeq The number of 
triangulations is known to be asymptotically bounded from above 
by an exponential function \cite{LectNotes}. This allows to 
rewrite the weighting function as \beq W(k_0,N_3,T) = 
f(k_0,N_3)e^{k_V N_3 s(\zeta)} \eeq where $f(k_0,N_3)$ is a 
function that grows sub-exponentially with $N_3$, and $s(\zeta)$ 
is some function that can (in principle) be found using 
combinatorics. As a result of this substitution, the partition 
function becomes \beq \label{Z3} Z(k_0,N_3)\sim\sum_{T} 
f(k_0,N_3)e^{k_V N_3(s(\zeta)+k_0 \zeta)}. \eeq 

Still working in the large $k_V N_3$ regime, the sum can be 
replaced by an integral; the integration variable can be taken to 
be $\zeta$ so that \beq \label{Z4} Z \sim 
\int_{\zeta_{min}}^{\zeta_{max}} f(k_0,N_3)e^{k_V 
N_3(s(\zeta)+k_0 \zeta)}\, d\zeta. \eeq The limits on the 
integral indicate the minimum and maximum values that the 
parameter $\zeta$ can take. The integral is dominated by the 
configurations for which the expression $s(\zeta)+k_0\zeta$ in 
the exponential is maximized in the allowed range 
$\zeta_{min}<\zeta<\zeta_{max}$. The position of the maximum 
depends on the precise form of $s(\zeta)$, but for $k_0$ large 
enough, it should occur at $\zeta_{max}$. Contributions to the 
partition function at other values of $\zeta$ are exponentially 
smaller, so the partition function can be simplified further to 
\beq Z \label{Z5} \sim \int_{\zeta_{min}}^{\zeta_{max}} 
f(k_0,N_3)e^{k_V N_3(s(\zeta)+k_0\zeta)} 
\delta(\zeta-\zeta_{max})\, d\zeta = f(k_0,N_3)e^{k_V 
N_3(s(\zeta_{max})+k_0\zeta_{max})}. \eeq The final result is 
that the macroscopic properties of space-time are determined by 
the triangulations with $\zeta=3.973$, or $\xi=5/4$; these 
configurations are formed by repetitively applying the $(2\ra 6)$ 
move in (\ref{oldmoves}).

As mentioned above, the importance of the Lorentzian model is 
that the upper value $5/4$ for $\xi$ is smaller than the 
analogous bound of $4/3$ in purely Euclidean triangulation models 
which are pathological \cite{LectNotes}. The correlation between 
the weaker upper bound for $\xi$ (and $\zeta$) in the Lorentzian 
model and the observation that the causal dynamical 
triangulations avoid the non-realistic features of non-causal 
models provides the hope that the Lorentzian model may describe 
at least some aspects of quantum gravity.

%%%%%%%%%%%%%%%%%%%%%%%%%%%%%%%%%%%%%%%%%%%%%%%%%%%%%%%%%%%%%%%%%%%%%%%%%%%%%%%%%%%%
\section{Variable Lapse \label{s_lapse}}

One of the important characteristics of the original Lorentzian 
model of dynamical triangulations is that all simplices of each 
type ($(3,1)$ or $(2,2)$) are exactly alike. In particular, all 
time-like edges have equal length - the model can be said to have 
a fixed `lapse.' In this section, the fixed lapse condition is 
relaxed without compromising the low-energy behavior of the 
model. 

This kind of generalization is known to be possible in $1+1$ 
dimensions \cite{FotiniLee}. There, the action is the sum of the 
areas of the simplices in the triangulation, \beq \label{S2d} 
S^{(2D)}=\lambda \sum_{i} A \eeq and the fixed lapse condition is 
equivalent to setting all the areas of the triangles to a 
particular value $A_0$. The variable lapse generalization 
consists of considering other triangulations in which the areas 
of the triangles are not all equal but where the average of these 
areas is still $A_0$. These configurations can be viewed as 
representing the same physical space-time as the fixed-lapse 
configuration, but implementing a different choice of foliation. 
Since the actions corresponding to the two types of 
configurations are equal, the low-energy behavior of the two 
models is the same. 

A similar argument can also be made in higher dimensions. 
Consider for example working with the action (\ref{Sgeneral}) and 
keeping the assumption that the simplices are either of the type 
$(3,1)$ or $(2,2)$. Instead of requiring that all the time-like 
edges have length-squared $\al$ as in the original model, suppose 
that these edge lengths can vary. In other words, tetrahedra may 
have all time-like edges of equal length or they may have some 
edges longer than others. The only requirement that has to be 
imposed is for the faces of neighboring tetrahedra to match so 
that they can be properly glued together into well-defined 
triangulation. This requirement always remains implicit and never 
appears in the equation of the defining action.

In this section, the simplices making up the space-time are 
labelled by an index $v$. Each simplex has a volume called $V_v$. 
The edges $e_j$ can be either space-like or time-like and each 
one may have a different length-squared. The types of simplices 
are distinguished by an index $i$ and the number of simplices of 
each type are denoted by $N^i_d$, with $d$ being the dimension. 
For example, $N_1^S$ is the number of space-like edges and 
$N_3^{i}$ is the number of tetrahedra of type $i$. The dihedral 
angle at an edge $e_j$ is written as $A(e_j)$ (note, however, 
that $A(e_j)$ is a function of the length of the edges 
neighboring $e_j$ as well as to the length of $e_j$ itself.)

The action for the variable-lapse model is of the form given by 
(\ref{Sgeneral}), \beq \label{S3new1} S_{new} =  -k_0 
\left(\sum_j 2\pi i V(e_j)\right)- \sum_v \left(k_0 \left( 
{\sum_{j}}^\prime V(e_{j})A(e_{j}) \right)  + \lambda V_v\right). 
\eeq The sum over $j$ in the first term is over all the edges, 
both space-like and time-like, in the triangulation. The second 
(primed) sum over $j$, however, is restricted to only those edges 
forming the skeleton of a particular simplex $v$. Some of the 
angles $A(e_j)$ are therefore subtended by space-like edges and 
some are subtended by time-like edges. Also, since space-like and 
time-like edges are treated together, the volumes $V(e_j)$ 
corresponding to time-like edges have absorbed factors of $i$; 
they can recovered by comparing this action to (\ref{Sgeneral}). 

There is no constraint in action (\ref{S3new1}) that forces all 
the edges to be of equal length, but whenever the simplices can 
be categorized into particular types, i.e. when several simplices 
have the same volumes and dihedral angles associated with them, 
the sums in the above formula can be simplified by grouping 
repeated terms together. There are at least two suggestive ways 
to rearrange the terms in the action. One way is to first group 
the terms arising from each simplex type together and then sum 
over the types, \beq \label{S3new2} S_{new} = -k_0 \left(\sum_j 
2\pi i V(e_j)\right) - \sum_i N^i_3 \left(\left( 
k_0{\sum_{j}}^\prime V^i(e_{j})A(e_{j})\right) + \lambda V^i 
\right). \eeq The second way is to exchange the order of 
addition. If the cosmological and curvature contributions are 
collected separately, then \beq \label{S3new3} S_{new} = -k_0 
\left(\sum_j 2\pi i V(e_j)\right) -k_0 \left( \sum_i N^i_3 
{\sum_{j}}^\prime V^i(e_{j})A(e_{j})\right) - \lambda \left( 
\sum_i N^i_3 V^i \right). \eeq 

Such a model model with variable lapse can be reduced to the 
original one under some conditions. There are several sets of 
constraints that are analogous to setting the average area of 
triangles to $A_0$ in the $1+1$ dimensional model (\ref{S2d}). 
For example, comparing (\ref{S3new3}) with (\ref{S3full}), it 
appears natural to set \beqa \label{newcond1} -k_0 \left(\sum_j 
2\pi i V(e_j)\right) &=&  k_0 \left( \frac{2\pi}{i}N_1^S + 
\sqrt{{\al_{eff}}} 2\pi N_1^T\right), 
\\ \label{newcond2} -k_0 \left( \sum_i N_3^i \sum_{j^\prime} 
V_i(e_{j^\prime})A(e_{j^\prime})\right) &=& - k_0 \bigg(
  \frac{2}{i}N_3^{(2,2)}\sin^{-1}\frac{-i2\sqrt{2}\sqrt{2{\al_{eff}}+1}}{4{\al_{eff}}+1} 
  +
  \frac{3}{i}N_3^{(3,1)}\cos^{-1}\frac{-i}{\sqrt{3}\sqrt{4{\al_{eff}}+1}} 
  \nonumber \\&&\qquad  +4N_3^{(2,2)}\cos^{-1}\frac{-1}{4{\al_{eff}}+1} +
   3N_3^{(3,1)}\cos^{-1}\frac{2{\al_{eff}}+1}{4{\al_{eff}}+1}\bigg),
\\ \label{newcond3} \lambda \left( 
\sum_i N_3^i V_i \right) &=& \lambda\left( 
N_3^{(2,2)}\frac{1}{12}\sqrt{4{\al_{eff}}+2} + 
N_3^{(3,1)}\frac{1}{12}\sqrt{3{\al_{eff}}+1}\right). \eeqa The 
expressions on the right hand side are terms from the action 
(\ref{S3full}) of the original model with an effective value 
${\al_{eff}}$ for the time-like edge-length. 

It is interesting that if the number of simplices in the variable 
and fixed lapse models are also set to be the same, i.e. \beq 
\label{newcond4} \sum_i N_3^i = N_3^{(3,1)} + N_3^{(2,2)}, \qquad 
\qquad \sum_j 1 = N_1^S + N_1^T \eeq where the sum over $j$ gives 
the total number of edges in the triangulation, then 
(\ref{newcond1})-(\ref{newcond3}) acquire geometrical 
interpretations. Equation (\ref{newcond3}), for instance, 
requires that the average volume of the simplices in the new 
model be the same as the volume of a simplex in the fixed-lapse 
model. Equations (\ref{newcond1}) and (\ref{newcond2}) require 
similar averages to hold for edge lengths and dihedral angles. 
Physically, such conditions would imply that, on the large scale, 
the variable-lapse and fixed-lapse models would be 
indistinguishable. One could also impose more stringent 
conditions such as requiring that the averaging conditions hold 
true separately on every slice of the triangulation. This would 
implement a form of time-symmetry that may simplify calculations 
of, for example, the correlation function between two slices in 
the space-time.

The form of the action appearing in (\ref{S3new2}) suggests a 
different set of averaging conditions: it is possible to have 
(\ref{newcond1}) together with the a new condition being the sum 
of (\ref{newcond2}) and (\ref{newcond3}). Adding these last two 
equation together effectively means that the averaging allows the 
mixing of the curvature and cosmological contributions to the 
action; the resulting condition would be weaker than 
(\ref{newcond1})-(\ref{newcond3}) separately. The effect of this 
mixing would only have noticeable consequences if an observer had 
the opportunity to probe the individual simplices of the 
triangulation.

Do these new models lie in the same equivalence class as the 
original fixed-lapse model? The answer to this question is in 
direct analogy with the discussion given in the 1+1 dimensional 
case \cite{FotiniLee}. Although the presence of new types of 
simplices implies that the model has new degrees of freedom, the 
averaging conditions for the new triangulations freeze some of 
these new degrees of freedom to the effect that the resulting 
action can be replaced by an effective fixed-lapse action. Just 
as in the discussion of the $1+1$ dimensional model, the possible 
arrangements of lapses in each triangulation or each slice in the 
triangulation are not summed over. Thus, even if a wider variety 
of configurations become available under these new conditions, 
the model stays in the same equivalence class as the fixed-lapse 
model, in the sense that the partition functions for the two 
models are equal. It follows that models with varying lapse have 
the same low energy properties as the original fixed-lapse 
version. 

Similar arguments can be applied in higher dimensions $d$ as 
well. Conditions analogous to (\ref{newcond1})-(\ref{newcond4}) 
would then be interpreted as averaging over volumes of 
$d$-simplices, or curvatures concentrated on the $d-2$ 
dimensional faces.

%%%%%%%%%%%%%%%%%%%%%%%%%%%%%%%%%%%%%%%%%%%%%%%%%%%%%%%%%%%%%%%%%%%%%%%%%%%%%%%%%%%%%%%%%%
\section{Foliations with Punctures \label{s_broken}}

\begin{figure}
\begin{center}
  \includegraphics[scale=1]{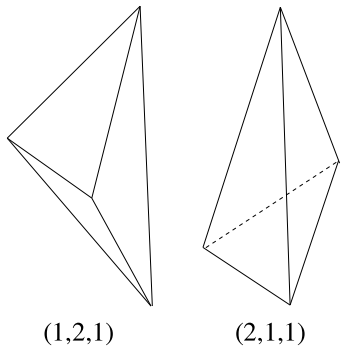}
  \end{center}
\caption{ \label{f_newsimplex} Simplices in 2+1 dimensions that 
span two foliation layers. }
\end{figure}

Another peculiar aspect of the original causal dynamical 
triangulation model is the arrangement of simplices in a layered 
structure. In this section, a new model is presented in which the 
foliation is `punctured' in the sense that a new type of simplex 
is allowed to probe multiple layers of tetrahedra. There are 
several types of tetrahedra that can puncture the foliation in 
this way: two of them are shown in figure \ref{f_newsimplex}. In 
the $(1,2,1)$ simplex, for example, two vertices are spatially 
separated from each other but are at the same time in the causal 
future of the third vertex and in the causal past of the fourth. 
The foliation, in the sense of the original triangulation model, 
is punctured by the $(1,2,1)$ tetrahedron because it does not 
have a space-like face that passes through the spatially 
separated vertices. Thus, when these simplices are present, it is 
difficult to construct a space-like hyper-surfaces that span all 
of the space-time.

The presence of (1,2,1) tetrahedra may affect the large volume 
behavior of dynamical triangulations. To investigate this 
possibility, consider a simple model in which there are $(3,1)$, 
$(2,2)$ and $(1,2,1)$ simplices, all space-like edges have 
length-squared equal to $1$, all time-like edges of the $(3,1)$ 
and $(2,2)$ tetrahedra and the shorter sides of the $(1,2,1)$ 
simplex have length-squared $-\al$, and the longer time-like 
edges of the $(1,2,1)$ tetrahedron have length-squared $-\beta$. 
Expressions for volumes and dihedral angles for the $(1,2,1)$ 
tetrahedron with this geometry can be computed using the methods 
of \cite{Hartle}. The volume is \beq V_{(1,2,1)} = 
\frac{1}{6}\sqrt{ 
 \frac{1}{4}\beta^2 -\frac{1}{4}\beta+\al\beta} 
\eeq and the dihedral angles $\theta_S, \theta_T$ and $\theta_L$ 
around the space-like, short time-like and long time-like edges, 
respectively, are \beq \cos \theta_S = 
\frac{4\alpha-2\beta+1}{4\alpha+1}, \qquad \cos \theta_T = 
\frac{\sqrt{\beta}}{\sqrt{4\alpha+1}\sqrt{4\beta +1}}, \qquad 
\cos \theta_L = \frac{4\alpha+2-\beta}{4\alpha-\beta}. \eeq 

The action for this model is an extension of $S_{original}$ in 
(\ref{S3full}) by terms specific to the new simplex type, \beqa S 
&= & S_{original} - k_0
   \left( \frac{1}{i}N_3^{(1,2,1)}\cos^{-1}\frac{(4\al-2\beta+1)}{4\al+1}
   \right) - \sqrt{\alpha} k_0
   \left( 4 N_3^{(1,2,1)} \cos^{-1}
            \frac{\sqrt{\beta}}{\sqrt{4\al+1}\sqrt{\beta-4\al}}
   \right)  \nonumber \\
  &&+\sqrt{\beta} k_0
   \left( 2\pi N_1^L -
          N_3^{(1,2,1)}\cos^{-1} \frac{4\al+2-\beta}{4\al-\beta}
   \right) 
  -\lambda\left(N_3^{(1,2,1)} \frac{1}{6}
        \sqrt{\frac{1}{4}\beta^2-\frac{1}{4}\beta+\al\beta}
            \right). 
\eeqa The variable $N_1^{L}$ counts the number of the `long' 
time-like edges that have length-squared $-\beta$.

To study the statistical properties of this model, the action 
should be Wick-rotated into Euclidean space. For the purposes of 
this paper, this merely means choosing negative values for 
$\alpha$ and $\beta$ such that the action becomes purely 
imaginary. For the part of the action depending on the $(3,1)$ 
and $(2,2)$ simplices, this can be done by taking $\alpha\ra -1$ 
as before. As for the part of the action specific to the new 
simplex, a suitable $\beta$ can be found such that $S\ra iS_E$ 
and $S_E$ is real. After evaluating the angles and volumes, the 
general form of the action is \beq S_E = \left( \left(k_A k_0 + 
\lambda k_V \right)N_3^{(3,1)+(2,2)} + \left(k_A^\prime k_0 + 
\lambda k_V^\prime \right)N_3^{(1,2,1)} - \left( k_1 N_1^{S+T} + 
k_1^\prime N_1^{L}\right)\right), \eeq where $k_1$, $k_A$, and 
$k_V$ are the same as in (\ref{k1k3defs}). The primed variables 
can be computed by fixing $\beta$. $S_E$ can be written in a form 
similar to (\ref{canonicalaction}) by factorizing the volume 
terms. The result is \beq S_E = V \left(\lambda - k_0 
\zeta^\prime \right) \eeq where \beq V= k_V N_3^{(3,1)+(2,2)} + 
k_V^\prime N_3^{(1,2,1)}\eeq and \beq \label{zetapdef} 
\zeta^\prime = \frac{k_1}{k_0}\frac{N_1^{S+T}}{V} - k_A 
\frac{N_3^{(3,1)+(2,2)}}{V} + 
\frac{k_1^\prime}{k_0}\frac{N_1^L}{V} - k_A^\prime 
\frac{N_3^{(1,2,1)}}{V}. \eeq Here $\zeta^\prime$ is a 
generalization of $\zeta$ in (\ref{xidef}).

Since the form of the action $S_E$ is the same as 
(\ref{canonicalaction}), the statistical mechanics of this new 
model is very similar to that discussed previously. The partition 
function for the new model, \beq \label{Z1p} Z^\prime(k_0,N_3) = 
\sum_{T} W^\prime(k_0,N_3,T)e^{-k_V 
N_3(\lambda-k_0\zeta^\prime)}, \eeq is like (\ref{Z1}) but with a 
larger number of variables summed over. As before, $W^\prime$ 
should be asymptotically bounded by an exponential in the 
large-volume limit. The sum can therefore be replaced by an 
integral over $\zeta$, \beq \label{Z4p} Z^\prime \sim 
\int_{\zeta^\prime_{min}}^{\zeta^\prime_{max}} f(k_0,N_3)e^{V 
(s(\zeta^\prime) + k_0 \zeta^\prime)} \, d\zeta^\prime. \eeq The 
dominant contribution to the integral comes from the values of 
$\zeta^\prime$ that maximize the exponent. Under the same 
assumptions as in section \ref{s_model}, one concludes that for 
$k_0$ large enough, the partition function should be dominated by 
maximal values of $\zeta^\prime$. Thus, in that approximation, 
\beq \label{Z5p} Z^\prime \sim 
\int_{\zeta^\prime_{min}}^{\zeta^\prime_{max}} f^\prime(k_0,N_3) 
e^{V (s(\zeta^\prime) + k_0 \zeta^\prime)} 
\delta(\zeta^\prime-\zeta^\prime_{max})\, d\zeta^\prime = 
f^\prime(k_0,N_3) 
e^{V(s(\zeta^\prime_{max})+k_0\zeta^\prime_{max})}, \eeq which is 
completely analogous to (\ref{Z5}).

What is left to explore is how $\zeta^\prime_{max}$ differs from 
$\zeta_{max}$. The governing assumption is that the new parameter 
$\zeta^\prime_{max}$ can serve as an indicator of large-volume 
behavior just like $\xi$ in the case of the original Euclidean 
and Lorentzian models. If this assumption is correct, and if the 
two values $\zeta_{max}$ and $\zeta_{max}^\prime$ are the same, 
then the partition functions of the models would differ only by a 
multiplicative factor and the physics of the two models would be 
similar. To investigate the range of $\zeta^\prime$, one can 
follow an analogous procedure to that used in section 
\ref{s_model} and explore how various moves affect configurations 
of tetrahedra.

\begin{figure}
\begin{center}
  \includegraphics[scale=1]{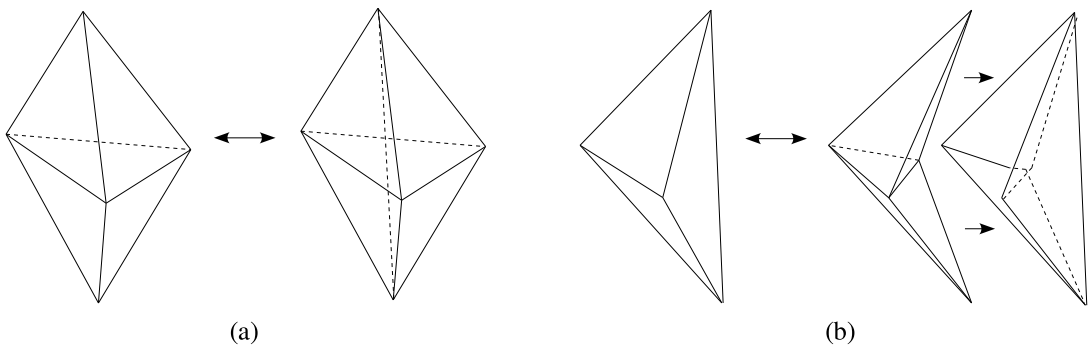}
  \end{center}
\caption{ \label{f_newmoves3} New moves manipulating $(1,2,1)$ 
simplices. (a) Two $(3,1)$ change into three $(1,2,1)$ 
tetrahedra. (b) One $(1,2,1)$ changes into two $(1,2,1)$ and two 
$(3,1)$ tetrahedra. The $(3,1)$ simplices appear distorted in the 
diagram.}
\end{figure}

Because $\zeta$ reduces to $\xi$ in the limit $N_3^{(1,2,1)}\ra 
0$, the moves defined in (\ref{oldmoves}) suggest that values for 
$\zeta^\prime$ within the range (\ref{zetabound}) should be 
allowed in the extended model as well. But new moves are now 
possible as well. In particular, it is important to consider 
moves that manipulate the number of $(1,2,1)$ tetrahedra. Two 
such moves are shown in figure \ref{f_newmoves3}. The first move 
is of limited applicability since it cannot be used repeatedly. 
It does, however, show how to introduce the foliation puncturing 
simplices into a triangulation that initially contains only 
$(3,1)$ and $(2,2)$ tetrahedra. The second move can be applied 
repeatedly and may therefore have an effect on the maximum value 
of $\zeta^\prime$. It is worth noting that other moves that 
rearrange simplices while keeping a boundary surface fixed are 
also possible. However, the two moves shown in the figure are 
sufficient to discuss the large-scale behavior of the extended 
model. 

The state of the configuration can be described by the vector 
\beq f^\prime = f^\prime (N_0, N_1^S, N_1^T, N_1^L,
                     N_3^{(3,1)}, N_3^{(2,2)}, N_3^{(1,2,1)}).
\eeq The two moves shown in the figure, denoted as 
$(2\ra3)^\prime$ and $(1\ra4)^\prime$, change this vector by \beqa
\Delta_{(2\ra3)^\prime} f^\prime &=& (0,\,0,\,0,\,1,\,-2,\,0,\,3), \nonumber \\
\Delta_{(1\ra4)^\prime} f^\prime &=& (1,\,2,\,2,\,0,\,2,\,0,\,1). 
\label{newmoves} \eeqa After a large number of moves of the 
$(1\ra 4)^\prime$ type, $\zeta^\prime$ would be \beq 
\label{zetapexp} \zeta^\prime = 
\frac{k_1}{k_0}\frac{4}{2k_V+k_V^\prime} - k_A 
\frac{2}{2k_V+k_V^\prime} - k_A^\prime \frac{1}{2k_V+k_V^\prime}. 
\eeq The numerical value of expression (\ref{zetapexp}) depends 
on the value of $\beta$, the length-squared of the long edge of 
the $(1,2,1)$ simplex, which should be between $-4$ and $-1$. 
Such values of $\beta$ give $\zeta^\prime$ in the range $ 6.42 < 
\zeta^\prime < 8.4.$ and take $\zeta^\prime_{max}$ above the 
Lorentzian level (\ref{zetabound}), but still keep it below the 
Euclidean level. Therefore, the new model should be expected to 
have a different low energy behavior than either the Euclidean 
model or the original Lorentzian model. 

Purely Euclidean models have the undesirable property that the 
most likely configurations have most of the tetrahedra connected 
to each other, failing to make up an extended space-time. (The 
Lorentzian models do not have this problem.) Since the new move 
(figure \ref{f_newmoves3} (b)) creates multiple tetrahedra that 
share the same edge, it is possible that the new Lorentzian model 
may suffer from a similar effect. It is not immediately clear, 
however, how strongly the pathological effect would be exhibited 
in the new model, and one may still hope that some of the 
desirable properties of the original Lorentzian model are 
preserved. It is safe to say that the behavior of the extended 
model should be intermediate between the purely Euclidean and the 
original Lorentzian cases. To make more precise statements about 
the low energy behavior, one would need to simulate the extended 
model on a computer.

As an aside, suppose that for some unknown reason, the frequency 
of the new simplex type relative to the old types is restricted. 
This condition would fix the number of moves (\ref{newmoves}) 
that would be allowed compared to the number of moves 
(\ref{oldmoves}). As a consequence, the last two terms in 
(\ref{zetapdef}) would be fixed, large scale triangulations would 
be generated by the old moves (\ref{oldmoves}), and 
$\zeta^\prime$ would not violate the Lorentzian bound 
(\ref{zetabound}). The triangulations arising from this model 
would look like configurations in the original model on large 
scales, but have small bubbles of $(1,2,1)$ tetrahedra on small 
scales; large clusters of foliation-puncturing tetrahedra would 
not be present. This kind of restriction on the partition 
function, however, is admittedly ad-hoc and at the moment does 
not have a physical or mathematical justification.

\section{Conclusion \label{s_conc}}

This paper investigated possible ways of relaxing assumptions 
related to the foliation structure in causal dynamical 
triangulation models in $2+1$ dimensions. Two results emerged 
from the discussion.

The first result is that the sizes and shapes of the simplices 
making up the triangulation can be made variable without 
compromising the statistical properties of the model. A way to 
understand this result is to say that when the configurations of 
simplices are constructed such that the average volumes or angles 
are the same as in the original fixed-lapse model, the action and 
therefore the partition function are unchanged. In this view, the 
requirement for all the simplices to be of the same size and 
shape is thus traded for a global constraint at the level of the 
partition function. The argument can be applied in higher 
dimensions as well.

The second result is concerned with the existence of space-like 
hyper-surfaces formed from tetrahedral faces. These 
hyper-surfaces, which appear naturally in the model with only 
$(3,1)$ and $(2,2)$ tetrahedra, can be punctured by introducing 
$(1,2,1)$ simplices. The partition function associated with this 
extended model is found to be different from the partition 
function of the original model, but it is unclear whether the 
extended model contains a similar pathology as that exhibited by 
purely Euclidean models. More precise statements about the 
large-scale properties of the proposed model (or other 
significant extensions of the original setup) would require 
detailed numerical simulations. It may be of some interest, 
however, that the possibility of a pathology can be removed if 
the sum over triangulations in the partition function is 
restricted in certain ways, for example by fixing the ratio of 
$(1,2,1)$ to the total number of simplices. In other words, 
allowing a small number of $(1,2,1)$ tetrahedra into a 
configuration should not spoil the large-scale properties of 
causal dynamical triangulations.

\vspace{0.3cm} {\bf Acknowledgements.$\;\;\;$}I would like to 
thank Fotini Markopoulou, Lee Smolin, Mohammad Ansari, and Jan 
Ambj\o rn for comments, encouragement and helpful discussions.

\end{document}